\documentclass[superscriptaddress,twocolumn,secnumarabic,
nobibnotes,aps,prd,nofootinbib]{revtex4}
\usepackage{graphicx}
\usepackage{epsf}
\usepackage{bm}
\usepackage{amsmath}
\usepackage{amsfonts}
\usepackage{amssymb}
\usepackage{epstopdf}
\usepackage{natbib}
\usepackage{color}
\usepackage{color}
\providecommand{\U}[1]{\protect\rule{.1in}{.1in}}
\newcommand{\be}{\begin{equation}}
\newcommand{\ee}{\end{equation}}

\newcommand{\mincir}{\raise
-3.truept\hbox{\rlap{\hbox{$\sim$}}\raise4.truept\hbox{$<$}\ }}
\newcommand{\magcir}{\raise
-3.truept\hbox{\rlap{\hbox{$\sim$}}\raise4.truept\hbox{$>$}\ }}

\begin{document}

\title{2021-$H_0$ Odyssey: Closed, Phantom and Interacting Dark Energy Cosmologies}

\author{Weiqiang Yang}
\email{d11102004@163.com}
\affiliation{Department of Physics, Liaoning Normal University, Dalian, 116029, P. R. China}

\author{Supriya Pan}
\email{supriya.maths@presiuniv.ac.in}
\affiliation{Department of Mathematics, Presidency University, 86/1 College Street, 
Kolkata 700073, India}

\author{Eleonora Di Valentino}
\email{eleonora.di-valentino@durham.ac.uk}
\affiliation{Institute for Particle Physics Phenomenology, Department of Physics, Durham University, Durham DH1 3LE, UK}

\author{Olga Mena}
\email{omena@ific.uv.es}
\affiliation{IFIC, Universidad de Valencia-CSIC, 46071, Valencia, Spain}

\author{Alessandro Melchiorri}
\email{alessandro.melchiorri@roma1.infn.it}
\affiliation{Physics Department and INFN, Universit\`a di Roma ``La Sapienza'', Ple Aldo Moro 2, 00185, Rome, Italy}

%%%%%%%%%%%%%%%%%%%%%%%%%%%%%%%%%%%%%%%%%%%%%%%%%%%%%%%%%
\begin{abstract}
 
 Up-to-date cosmological data analyses have shown that \textit{(a)} a closed universe is preferred by the Planck data at more than $99\%$ CL, and \textit{(b)} interacting scenarios offer a very compelling solution to the Hubble constant tension.  In light of these two recent appealing scenarios, we consider here an interacting dark matter-dark energy model with a non-zero spatial curvature component and a freely varying dark energy equation of state in both the quintessential and phantom regimes. When considering Cosmic Microwave Background data only, a phantom and closed universe can perfectly alleviate the Hubble tension, without the necessity of a coupling among the dark sectors. Accounting for other possible cosmological observations compromises the viability of this very attractive scenario as a global solution to current cosmological tensions, either by spoiling its effectiveness concerning the $H_0$ problem, as in the case of Supernovae Ia data, or by introducing a strong disagreement in the preferred value of the spatial curvature, as in the case of Baryon Acoustic Oscillations.

\end{abstract}

\maketitle
%%%%%%%%%%%%%%%%%%%%%%%%%%%%%%%%%%%%%%%%%%%%%%%%%%%%%%%%%%%%%%%%%%

%%%%%%%%%%%%%%%%%%%%%%%%%%%%%%%%%%%%%%%%%%%%%%%%%%%%%%%%%%%%%%%%%%
\section{Introduction}

With the development of observational cosmology, our Universe is becoming more complex and mysterious. The dark energy issue is already an unsolved problem in modern cosmology and some recent observational evidences claim that our Universe is closed, despite its widely-believed flatness nature. Analyses of the latest Cosmic Microwave Background (CMB) measurements from the Planck 2018 legacy release with the baseline \texttt{Plik} likelihood point towards the possibility of a closed Universe at more than three standard deviations~\cite{Aghanim:2018eyx,Handley:2019tkm,DiValentino:2019qzk,DiValentino:2020hov}. This observational outcome, undoubtedly, is one of the most significant results at present times, and puts a question mark on the standard $\Lambda$CDM scenario and the inflationary prediction of a flat universe. Even though a flat universe cannot be excluded, accordingly to other complementary results performed with the \texttt{CamSpec} alternative likelihood~\cite{Efstathiou:2020wem,Efstathiou:2019mdh}, the obtained marginalized constraints using this alternative likelihood still prefer a closed model at a higher significance level, above $99 \%$ CL, see Ref.~\cite{Efstathiou:2019mdh}. 

Such a closed universe has been found to exacerbate previous tensions in some cosmological parameters. For example, the existing $4.4\sigma$ tension on the Hubble constant value between Planck CMB data~\cite{Aghanim:2018eyx} (within the minimal $\Lambda$CDM model) and the SH0ES collaboration~\cite{Riess:2019cxk} (R19) is increased to the $5.4\sigma$~\cite{DiValentino:2019qzk} level\footnote{See also~\cite{DiValentino:2020zio,DiValentino:2020srs,DiValentino:2021izs} for a recent overview.}. Additionally, if the curvature parameter is allowed to freely vary, there is an increase in the tensions between Planck CMB and Baryon Acoustic Oscillation (BAO) observations~\cite{DiValentino:2019qzk,Handley:2019tkm} and between Planck CMB and the full-shape galaxy power spectrum measurements~\cite{Vagnozzi:2020zrh}. 

Therefore, the evidence for a closed universe has raised some unavoidable questions that need to be answered. A partial resolution of these problems  can be achieved by the introduction a phantom dark energy component~\cite{DiValentino:2020hov}, where the discrepancy in the $H_0$ measurements between Planck and SH0ES collaborations can be solved in the closed universe model, but the tension with the BAO data still persists. Consequently, possible solutions may reside on the dark sector microphysics. In this regard, looking for some physically motivated dark energy models that could lead to an effective phantom dark energy equation of state in a closed universe model may provide the key to fully address the current cosmological tensions. Interacting scenarios originally proposed to explain the cosmological constant first \cite{Wetterich:1994bg} and coincidence problem later \cite{Amendola:1999er} 
got serious attention since the end of 90's with some appealing outcomes \cite{Cai:2004dk,Barrow:2006hia,delCampo:2008sr,Amendola:2011ie,Pettorino:2012ts,Salvatelli:2014zta,Wang:2014xca,Tamanini:2015iia,Boehmer:2015kta,Boehmer:2015sha,Casas:2015qpa,Pourtsidou:2016ico,Nunes:2016dlj,Pan:2016ngu,Sharov:2017iue,Yang:2017yme,Yang:2017ccc,Yang:2017zjs,An:2017crg,Mifsud:2017fsy,vandeBruck:2017idm,Cardenas:2018qcg,vonMarttens:2018iav,Yang:2018qec,Cruz:2019kgz,Paliathanasis:2019hbi,Barrow:2019jlm,Cheng:2019bkh,Benetti:2019lxu,Kase:2019mox,vonMarttens:2020apn,Pan:2020mst,Hogg:2020rdp,Lucca:2020zjb,Rezaei:2020yhr,Jimenez:2020npm} 
(see also two reviews~\cite{Bolotin:2013jpa,Wang:2016lxa}). Indeed, it was pointed out that when there is a coupling between the dark matter and the dark energy sectors, the system may naturally resemble an effective phantom $w_x <-1$ cosmology~\cite{Huey:2004qv,Wang:2005jx,Das:2005yj,Sadjadi:2006qb,Pan:2014afa}.

Motivated by this interesting possibility,  we consider here a generalized cosmic scenario, namely, the interacting dark matter (DM)-dark energy (DE) scenario within the context of a non-flat model of our Universe where the dark energy enjoys a constant equation of state other than $w_x =-1$. Recently, interacting dark sectors  have received plenty of attention in the literature due to their effectiveness in alleviating the Hubble constant tension  (see Section 8 of the Review~\cite{DiValentino:2021izs}, and in particular~\cite{Kumar:2017dnp,DiValentino:2017iww,Yang:2018euj,Yang:2018uae,Kumar:2019wfs,DiValentino:2019ffd,DiValentino:2019jae,Pan:2019jqh,Pan:2019gop,Pan:2020bur,Gomez-Valent:2020mqn,Lucca:2020zjb,Yang:2020uga,Yang:2019uog,Yang:2018ubt}).  Here we extend our recent work~\cite{DiValentino:2020kpf} (where the equation of state of DE was fixed at $w_x = -0.999$, that means, it effectively mimics the vacuum energy)  
by allowing the dark energy equation of state to freely vary in the quintessence and phantom regimes. 
Therefore, the present article is a more generalized  cosmological scenario which aims to investigate a number of open issues, namely, the curvedness of the universe (see Ref. \cite{Dossett:2012kd}, an earlier article in this direction), the possibility of an interaction in the dark sector, the existing cosmological tensions and finally the nature of the dark energy equation of state. So far we are aware of the recent literature, this is the first time we are reporting the aforementioned issues in a single work. It is important to note that, as the main recipe of this work is the non-flat universe which eventually exacerbates the tensions in the cosmological parameters, see e.g. \cite{DiValentino:2019qzk}, therefore, it is essential to understand whether there exists any cosmological scenario that can effectively alleviate or solve these tensions irrespective of the curved geometry of the universe. According to the existing literature, most of the investigations headed in this direction assumed the spatial flatness of the universe.      

The article has been organized as follows. Section~\ref{sec-ide} describes the basic equations of a general interacting scenario assuming a non-flat background of our Universe. Section~\ref{sec-data} presents the observational data that are used to constrain the model. In Sec.~\ref{sec-results} we discuss the constraints on the scenario explored here and finally in Sec.~\ref{sec-conclusion} we briefly summarize the main conclusions.

\section{Interacting dark energy in a curved Universe}
\label{sec-ide}

Assuming a Friedmann-Lema\^{i}tre-Robertson-Walker (FLRW) universe with a non-zero spatial curvature, we consider a generalized cosmological scenario where the dark fluids of the universe, namely the pressureless (or cold) DM and the DE interact with each other by allowing a transfer of energy and/or momentum between them. The other fluids of the universe, instead, such as radiation and baryons, do not take part in the interaction process. The energy densities of the pressureless DM and DE are respectively denoted by $\rho_c$ and $\rho_x$ and their corresponding pressure components are $p_c$, which is zero, and $p_x$. We further assume that DE has a constant equation-of-state parameter $w_x = p_x/\rho_x$ different from $w_x = -1$. The conservation equations of the DM and DE fluids at the background level are coupled via an arbitrary function, known as the interacting function $Q$: 
\begin{eqnarray}
\label{eq:continuitydensityc}
\dot{\rho}_c+3{\cal H}\rho_c &=& -Q\,, \\
\label{eq:continuitydensityx}
\dot{\rho}_x+3{\cal H}(1+w)\rho_x &=&+Q\,,
\end{eqnarray}
where an overhead dot denotes the derivative with respect to the conformal time $\tau$, and ${\cal H} \equiv \dot{a}/a$ is the conformal Hubble rate. The sign of the interaction function $Q$ determines the direction of the energy and/or momentum flow. Here, $Q > 0$ denotes the transfer of energy and/or momentum from pressureless DM to DE, while $Q <0$ describes exactly the opposite scenario. As already mentioned, apart from the dark fluids, none of the remaining fluids take part in the interaction and therefore they obey the standard conservation laws. Once a specific interaction function $Q$ is prescribed, then, by solving the conservation equations Eqs.~(\ref{eq:continuitydensityc}) and (\ref{eq:continuitydensityx}), either analytically or numerically, together with the equation for the Hubble rate evolution, it is possible to know the evolution of the universe. Thus, the choice of the interaction function is essential to determine the dynamical evolution of the universe. In the present paper we shall work with a very well known form for the interaction function~\cite{He:2008si,Valiviita:2008iv,Gavela:2009cy,Gavela:2010tm,Honorez:2010rr}:    

\begin{eqnarray}
Q = 3  \xi {\cal H} \rho_x\,,
\label{eq:coupling}
\end{eqnarray}
where $\xi$ is a dimensionless coupling parameter.
Apart from the modifications at the background level, the presence of the coupling also affects the equations at the level of perturbations, which, in the synchronous gauge are given by the following expressions~\cite{Valiviita:2008iv,Gavela:2009cy,Gavela:2010tm}:
\begin{eqnarray}
\label{eq:deltac}
\dot{\delta}_c &=& -\theta_c - \frac{1}{2}\dot{h} +\xi{\cal H}\frac{\rho_x}{\rho_c}(\delta_x-\delta_c)+\xi\frac{\rho_x}{\rho_c} \left ( \frac{kv_T}
{3}+\frac{\dot{h}}{6} \right )\,, \\
\label{eq:thetac}
\dot{\theta}_c &=& -{\cal H}\theta_c\,,\\
\label{eq:deltax}
\dot{\delta}_x &=& -(1+w) \left ( \theta_x+\frac{\dot{h}}{2} \right )-\xi \left ( \frac{kv_T}{3}+\frac{\dot{h}}{6} \right ) \nonumber \\
&&-3{\cal H}(1-w) \left [ \delta_x+\frac{{\cal H}\theta_x}{k^2} \left (3(1+w)+\xi \right ) \right ]\,,\\
\label{eq:thetax}
\dot{\theta}_x &=& 2{\cal H}\theta_x+\frac{k^2}{1+w}\delta_x+2{\cal H}\frac{\xi}{1+w}\theta_x-\xi{\cal H}\frac{\theta_c}{1+w}\,.
\end{eqnarray}
In the equations above, $\delta_c$ and $\delta_x$ denote the density perturbations for the DM and DE fluids respectively, $\theta_c$ and $\theta_x$ refer to the DM and DE velocity divergences respectively, $v_T$ is the center of mass velocity for the total fluid and $h$ is the synchronous gauge metric perturbation. Note that the sound speed for the DE fluid has been fixed to $1$. For issues concerning the required initial conditions for the DE coupled fluid perturbations, see Refs.~\cite{Gavela:2010tm,Majerotto:2009np} for a full discussion of the adiabaticity of the initial conditions (see also Refs.~\cite{Doran:2003xq,Ballesteros:2010ks}). 
We follow here the prescription derived in Ref.~\cite{Gavela:2010tm} providing the initial conditions for the DE density contrast and velocity divergence: 
\begin{eqnarray}
\delta_x^{\rm in}(\eta) &=& \frac{3}{2}\frac{(2\xi-1-w)(1+w+\xi/3)}{12w^2-2w-3w\xi+7\xi-14}\delta_{\gamma}^{\rm in}(\eta)\,,\\
\theta_x^{\rm in}(x) &=& \frac{3}{2}\frac{k\eta(1+w+\xi/3)}{2w+3w\xi+14-12w^2-7\xi}\delta_{\gamma}^{\rm in}(\eta)\,,
\label{eq:initialconditions}
\end{eqnarray}
where $\eta= k \tau$ and $\delta_{\gamma}^{\rm in}(\eta)$ is the initial conditions for the photon density perturbations.

Plenty of work has been devoted to establish the stability conditions for their evolution in cosmic time. 
A word of caution is needed here. It has been shown that interacting dark energy schemes may be subject to instabilities in the evolution of the perturbations~\cite{Valiviita:2008iv,He:2008si}. In Ref.~\cite{Gavela:2009cy} it was stated that as long as the so-called \emph{doom factor} 
\begin{equation}
\textbf{d}= \frac{Q}{3\mathcal{H}(1+w) \rho_x}
\end{equation}
is negative, the evolution of the system will be free of instabilities. Therefore, $(1+w)$ and $\xi$ must have opposite signs, accordingly to Ref.~\cite{Gavela:2009cy}. Notice however that our definition of the exchange rate differs from a minus sign from that of the reference above and consequently this stability condition for our case is translated into a stable parameter space in which $(1+w)$ and $\xi$ must have the same sign. Therefore, in the phantom regime in which $(1+w)$ is a negative quantity the dimensionless coupling $\xi$ must be also negative. On the other hand, in the quintessence region $\xi$ must be positive. For early work on instabilities in coupled models, see Refs.~\cite{Valiviita:2008iv,He:2008si,Jackson:2009mz, Gavela:2010tm,Clemson:2011an}. For more recent analyses which have also considered this issue, see Refs.~\cite{Li:2014eha,Li:2014cee,Guo:2017hea,Zhang:2017ize,Guo:2018gyo,Yang:2018euj,Dai:2019vif}.

While the interaction function in Eq.~(\ref{eq:coupling}) was initially motivated from a pure phenomenological perspective, a recent investigation shows that using a multi-scalar field action~\cite{Pan:2020zza}, the coupling function (\ref{eq:coupling}) can be derived\footnote{Additionally, the scalar field theory is not the only motivation to generate interaction functions of the form considered here. Another detailed investigation in Ref.~\cite{Pan:2020mst} shows that some existing cosmological theories with a Lagrangian description can also return such interaction functions.}. Therefore, the interaction model of the form given by Eq.~(\ref{eq:coupling}) also benefits from a solid theoretical motivation following some action principle.

\section{Observational data and methodology}
\label{sec-data}

In this section we discuss the observational data and the statistical methodology that we use to constrain the interacting scenarios of our interest. In what follows we describe the main observational data sets: 

\begin{itemize}

\item {\bf Planck 2018 CMB data}: we consider the Cosmic Microwave Background (CMB) measurements from the Planck 2018 legacy release, precisely the CMB temperature and polarization angular power spectra {\it plikTTTEEE+lowl+lowE} \cite{Aghanim:2018eyx,Aghanim:2019ame}.

\item {\bf CMB lensing}: we add the Planck 2018 CMB lensing likelihood~\cite{Aghanim:2018oex}, obtained from measurements of the trispectrum.
    
\item {\bf BAO}: various measurements of the Baryon Acoustic Oscillations (BAO) from different galaxy surveys, such as 6dFGS~\cite{Beutler:2011hx}, SDSS-MGS~\cite{Ross:2014qpa}, and BOSS DR12~\cite{Alam:2016hwk}, as used by the Planck collaboration~\cite{Aghanim:2018eyx}, have been used. 
    
\item {\bf Pantheon}: we include the Pantheon sample of the Supernovae Type Ia distributed in the redshift interval $z \in [0.01, 2.3]$~\cite{Scolnic:2017caz}.
   
\item {\bf R19}: we also include the measurement of the Hubble constant by the SH0ES collaboration in 2019~\cite{Riess:2019cxk}, yielding $H_0 = 74.03 \pm 1.42$ km/s/Mpc at $68\%$ CL.

\item {\bf DES}: we have also considered the galaxy clustering and cosmic shear measurements from the Dark Energy Survey (DES) combined-probe Year 1 results~\cite{Troxel:2017xyo, Abbott:2017wau, Krause:2017ekm}, as used by the Planck 2018 collaboration in~\cite{Aghanim:2018eyx}.

\end{itemize}

To constrain this interacting scenario, we have used a modified version of the well known cosmological package \texttt{CosmoMC}~\cite{Lewis:2002ah,Lewis:1999bs}, which is publicly available~\footnote{\url{http://cosmologist.info/cosmomc/}}. This modified code has been previously used in a number of published works in the literature~\cite{Gavela:2010tm,Salvatelli:2013wra,DiValentino:2017iww,DiValentino:2019jae,DiValentino:2019ffd,DiValentino:2020leo,DiValentino:2020kpf}. The \texttt{CosmoMC} package is equipped with a convergence diagnostic based on the Gelman-Rubin criterion~\cite{Gelman:1992zz} and supports the Planck 2018 likelihood~\cite{Aghanim:2019ame}. The flat priors on the free parameters of the scenario explored here are displayed in Tab.~\ref{priors}. We also present here Bayesian evidence analyses in order to assess the performance of the interacting scenarios compared to the spatially flat $\Lambda$CDM scenario. The computation of Bayesian evidences in terms of the Bayes factors $\ln B_{ij}$ (here $i$ and $j$ refer to the underlying models $M_i$ and $M_j$, in which one is the reference model) is straightforward once the MCMC chains are obtained. We make use of the publicly available package \texttt{MCEvidence}\footnote{ \href{https://github.com/yabebalFantaye/MCEvidence}{github.com/yabebalFantaye/MCEvidence}~\cite{Heavens:2017hkr,Heavens:2017afc}.} to compute the Bayesian evidences. 
The numerical values of $\ln B_{ij}$ provide the statistical support for the models with  respect to the reference model, classifying it by the revised Jeffreys scale by Kass and Raftery as given in  Ref.~\cite{Kass:1995loi,Trotta:2008qt}: (i) for $0 \leq | \ln B_{ij}|  < 1$, the model has an inconclusive evidence, (ii) for $1 \leq | \ln B_{ij}|  < 2.5$, the model has a weak evidence, (iii) for $2.5 \leq | \ln B_{ij}|  < 5$, the model has a moderate evidence, and, finally, (iv)  for $ | \ln B_{ij} | \geq 5$, the model has a strong evidence. 

\begin{table}
\begin{center}
\begin{tabular}{|c|c|c|c|c|}
\hline
Parameter                    & prior (phantom) & prior (quintessence) \\
\hline
$\Omega_{\rm b} h^2$         & $[0.005,0.1]$ & $[0.005,0.1]$ \\
$\Omega_{\rm c} h^2$         & $[0.001,0.99]$ & $[0.001,0.99]$\\
$100\theta_{MC}$             & $[0.5,10]$ & $[0.5,10]$\\
$\tau$                       & $[0.01,0.8]$ & $[0.01,0.8]$\\
$n_\mathrm{S}$               & $[0.7,1.3]$ & $[0.7,1.3]$\\
$\log[10^{10}A_{s}]$         & $[1.7, 5.0]$ & $[1.7, 5.0]$\\
$\Omega_{K0}$                & $[-2,2]$   & $[-2,2]$ \\
$\xi$                        & $[-1,0]$   & $[0, 1]$\\
$w_x$                        & $[-3, -1]$ & $[-1, 0]$\\
\hline %S
\end{tabular}
\end{center}
\caption{Flat priors on the main cosmological parameters used in this work.}
\label{priors}
\end{table}

\section{Observational Results}
\label{sec-results}
The stability criteria for interacting dark sector models governed by the interaction function given by Eq.~(\ref{eq:coupling}) requires to analyze separately the two allowed regions, namely, \textit{(i)} $\xi < 0$, $w_x < -1$, named as \textbf{IDEp}, and \textit{(ii)} $\xi >0$, $w_x> -1$, referred to as \textbf{IDEq}. In the following we shall describe the results arising from the analyses of the different observational data sets considered here within these two regimes.

%%%%%%%%%%%%%%%%%%%%%%%%%%%%%%%%%%%%%%%%%%%%

\begingroup                                                                                                                   \squeezetable                                        
\begin{center}                                                                                                                 
\begin{table*}                                                                                                                 \resizebox{\textwidth}{!}{   
\begin{tabular}{cccccccc}                                                                                               
\hline\hline                                                                                                                    
Parameters & CMB & CMB+lensing & CMB+Pantheon & CMB+R19 & CMB+DES \\ \hline

$\Omega_{\rm c} h^2$ & $    0.132_{-    0.014-    0.018}^{+    0.008+    0.021}$ & $    0.133_{-    0.013-    0.016}^{+    0.008+    0.018}$  & $    0.133_{-    0.014-    0.018}^{+    0.009+    0.020}$ & $    0.134_{-    0.015-    0.018}^{+    0.007+    0.023}$ & $    0.1226_{-    0.0048-    0.0067}^{+    0.0022+    0.0088}$  \\

$\Omega_{\rm b} h^2$ & $    0.02264_{-    0.00017-    0.00033}^{+    0.00018+    0.00032}$ & $    0.02250_{-    0.00016-    0.00031}^{+    0.00016+    0.00031}$  & $    0.02257_{-    0.00016-    0.00036}^{+    0.00019+    0.00031}$ & $    0.02261_{-    0.00017-    0.00033}^{+    0.00017+    0.00034}$   & $    0.02248_{-    0.00016-    0.00030}^{+    0.00016+    0.00031}$   \\

$100\theta_{MC}$ & $    1.04042_{-    0.00058-    0.0013}^{+    0.00078+    0.0012}$ & $    1.04030_{-    0.00058-    0.0011}^{+    0.00066+    0.0010}$  & $    1.04037_{-    0.00063-    0.0012}^{+    0.00071+    0.0011}$ & $    1.04035_{-    0.00057-    0.0013}^{+    0.00078+    0.0012}$ & $    1.04080_{-    0.00037-    0.00078}^{+    0.00042+    0.00073}$ \\

$\tau$ & $    0.0478_{-    0.0077-    0.017}^{+    0.0079+    0.016}$ & $    0.0494_{-    0.0077-    0.017}^{+    0.0085+    0.016}$  & $    0.0493_{-    0.0074-    0.017}^{+    0.0077+    0.016}$  & $    0.0479_{-    0.0077-    0.017}^{+    0.0088+    0.017}$  & $    0.0542_{-    0.0076-    0.015}^{+    0.0075+    0.016}$  \\

$n_s$ & $    0.9716_{-    0.0052-    0.011}^{+    0.0051+    0.010}$ & $    0.9690_{-    0.0046-    0.0091}^{+    0.0046+    0.0092}$  & $    0.9701_{-    0.0047-    0.0094}^{+    0.0055+    0.0093}$ & $    0.9709_{-    0.0047-    0.0093}^{+    0.0048+    0.0091}$  & $    0.9679_{-    0.0046-    0.0088}^{+    0.0045+    0.0092}$ \\

${\rm{ln}}(10^{10} A_s)$ & $    3.026_{-    0.018-    0.036}^{+    0.017+    0.035}$ & $    3.029_{-    0.016-    0.037}^{+    0.017+    0.033}$   & $    3.030_{-    0.016-    0.035}^{+    0.016+    0.037}$ & $    3.027_{-    0.017-    0.036}^{+    0.017+    0.036}$  & $    3.040_{-    0.016-    0.033}^{+    0.016+    0.032}$ \\

$w_x$ & $   >-1.90\,>-2.70$ & $   >-1.61\,>-2.0$  & $   -1.31_{-    0.10-    0.24}^{+    0.14+    0.22}$ & $   -2.03_{-    0.30-    0.73}^{+    0.43+    0.67}$ & $ > -1.57 > -1.79 $  \\

$\xi$ & $   >-0.108\,>-0.229$ & $   >-0.065\,>-0.111$  & $   >-0.102\,>-0.186 $ & $   >-0.099\,>-0.196 $ & $ > -0.012 > -0.030 $  \\ 

$\Omega_{K0}$ & $   -0.032_{-    0.012-    0.031}^{+    0.019+    0.028}$ & $   -0.0058_{-    0.0037-    0.012}^{+    0.0068+    0.010}$  & $   -0.028_{-    0.010 - 0.024}^{+    0.013+    0.020}$  & $   -0.0207_{-    0.0074 - 0.012}^{+    0.0055+    0.013}$ & $    0.0017_{-    0.0036-    0.0060}^{+    0.0031+    0.0064}$  \\

$\Omega_{m0}$ & $  0.39_{-    0.15-    0.24}^{+    0.12+    0.24}$ & $    0.27_{-    0.11-    0.14}^{+    0.06+    0.15}$   & $    0.419_{-    0.042-    0.072}^{+    0.037+    0.074}$ & $    0.288_{-    0.029-    0.041}^{+    0.018+    0.048}$  & $    0.219_{-    0.036-    0.071}^{+    0.035+    0.068}$  \\

$\sigma_8$ & $    0.82_{-    0.12-    0.19}^{+    0.08+    0.21}$ & $    0.83_{-    0.11-    0.16}^{+    0.07+    0.18}$   & $    0.769_{-    0.040-    0.081}^{+    0.054+    0.076}$ & $    0.880_{-    0.061-    0.11}^{+    0.070+    0.11}$ & $    0.898_{-    0.068-    0.115}^{+    0.054+    0.123}$ \\

$H_0$ & $   66_{-   15-   19}^{+    7+   24}$ & $   77_{-    16-    18}^{+    9+    21}$  & $   61.1_{-    2.5-    5.4}^{+    2.7+    5.0}$ & $   73.9_{-    1.4-    2.7}^{+    1.4+    2.8}$  & $   82.3_{-    8.0-   12}^{+    5.4+   14}$ \\

$S_8$ & $    0.908_{-    0.061-    0.13}^{+    0.077+    0.12}$ &  $    0.782_{-    0.048-    0.087}^{+    0.052+    0.084}$  & $    0.906_{-    0.054-    0.081}^{+    0.040+    0.098}$ & $    0.859_{-    0.039-    0.065}^{+    0.036+    0.069}$  & $    0.761_{-    0.021-    0.046}^{+    0.026+    0.041}$  \\

$r_{\rm{drag}}$ & $  147.38_{-    0.32-    0.62}^{+    0.35+    0.62}$ &  $  147.37_{-    0.30-    0.62}^{+    0.31+    0.61}$  & $  147.32_{-    0.30-    0.63}^{+    0.35+    0.58}$  & $  147.36_{-    0.30-    0.60}^{+    0.30+    0.59}$  & $  147.32_{-    0.30-    0.63}^{+    0.30+    0.59}$  \\

%\hline
%$\chi^2_{\rm bf}$ & $2761.95$ & $2778.07$ & $3799.24$ & $2765.18$ \\
%$\Delta \chi^2_{\rm bf}$ & $-10.7$ & $-3.97$ & $-8.26$ & $-26.66$ \\

\hline\hline                                               
\end{tabular}          }                                     
\caption{68\% and 95\% CL constraints on several free and derived parameters of the interacting scenario \textbf{IDEp} ensuring the stability criterion using a variety of cosmological data sets. }
\label{tab:IDEpOK}                                              
\end{table*}                                            
\end{center}                                                    
\endgroup

%%%%%%%%%%%%%%%%%%%%%%%%%%%%%%%%%%%%%%%%%%%%

%%%%%%%%%%%%%%%%%%%%%%%%%%%%%%%%%%%%%%%%%%%%

\begingroup                                                                                                                   \squeezetable                                        
\begin{center}                                                                                                                 
\begin{table*}                                                                                                                 \resizebox{\textwidth}{!}{   
\begin{tabular}{cccccccc}                                                                                               
\hline\hline                                                                                                                    
Parameters  & CMB+BAO & CMB+BAO & CMB+BAO & CMB+BAO\\
 && +Pantheon & +Pantheon+R19 & +Pantheon+R19+lensing\\\hline

$\Omega_{\rm c} h^2$  & $    0.135_{-    0.010-    0.015}^{+    0.010+    0.015}$   & $    0.1350_{-    0.0098-    0.014}^{+    0.0097+    0.014}$  & $    0.135_{-    0.012-    0.015}^{+    0.008+    0.016}$ & $    0.134_{-    0.012-    0.015}^{+    0.008+    0.016}$ \\

$\Omega_{\rm b} h^2$ & $    0.02240_{-    0.00016-    0.00031}^{+    0.00016+    0.00032}$   & $    0.02240_{-    0.00015-    0.00030}^{+    0.00016+    0.00031}$ & $    0.02239_{-    0.00016-    0.00029}^{+    0.00015+    0.00030}$ & $    0.02241_{-    0.00015-    0.00030}^{+    0.00015+    0.00030}$  \\

$100\theta_{MC}$  & $    1.04014_{-    0.00058-    0.0010}^{+    0.00057+    0.0011}$   & $    1.04013_{-    0.00061-    0.0010}^{+    0.00055+    0.0011}$ & $    1.04018_{-    0.00054-    0.0011}^{+    0.00060+    0.0010}$ & $    1.04017_{-    0.00058-    0.0011}^{+    0.00059+    0.0010}$ \\

$\tau$ & $    0.0547_{-    0.0083-    0.016}^{+    0.0073+    0.016}$  & $    0.0547_{-    0.0073-    0.015}^{+    0.0074+    0.015}$  & $    0.0544_{-    0.0078-    0.017}^{+    0.0079+    0.016}$ & $    0.0544_{-    0.0072-    0.014}^{+    0.0071+    0.018}$  \\

$n_s$  & $    0.9661_{-    0.0044-    0.0091}^{+    0.0045+    0.0092}$   & $    0.9660_{-    0.0047-    0.0085}^{+    0.0043+    0.0088}$ & $    0.9651_{-    0.0043-    0.0086}^{+    0.0044+    0.0091}$ & $    0.9655_{-    0.0043-    0.0085}^{+    0.0043+    0.0085}$ \\

${\rm{ln}}(10^{10} A_s)$  & $    3.044_{-    0.017-    0.031}^{+    0.015+    0.034}$   & $    3.044_{-    0.016-    0.031}^{+    0.015+    0.032}$ & $    3.045_{-    0.016-    0.031}^{+    0.016+    0.033}$ & $    3.044_{-    0.014-    0.028}^{+    0.014+    0.029}$ \\

$w_x$  & $   -1.12_{-    0.03}^{+    0.11}\,>-1.26$   & $   -1.079_{-    0.038}^{+    0.053}\,>-1.15$ & $   -1.113_{-    0.044-    0.091}^{+    0.045+    0.092}$ & $   -1.113_{-    0.044-    0.089}^{+    0.049+    0.084}$ \\

$\xi$  & $   -0.052_{-    0.025}^{+    0.045}\,>-0.101$   & $   -0.052_{-    0.024}^{+    0.043}\,>-0.10$ & $   >-0.062\,>-0.098 $ & $   >-0.065\,>-0.098 $ \\

$\Omega_{K0}$ & $   -0.0008_{-    0.0027-    0.0052}^{+    0.0026+    0.0054}$  & $   0.0000_{-    0.0022 - 0.0043}^{+    0.0022+    0.0044}$  & $   0.0008_{-    0.0021 - 0.0040}^{+    0.0020+    0.0041}$ & $   0.0008_{-    0.0021 - 0.0040}^{+    0.0021+    0.0043}$  \\

$\Omega_{m0}$  & $    0.333_{-    0.028-    0.045}^{+    0.025+    0.046}$   & $    0.339_{-    0.023-    0.037}^{+    0.022+    0.038}$  & $    0.323_{-    0.022-    0.034}^{+    0.017+    0.036}$ & $    0.323_{-    0.024-    0.034}^{+    0.019+    0.036}$  \\

$\sigma_8$  & $    0.764_{-    0.049-    0.075}^{+    0.040+    0.077}$    & $    0.756_{-    0.043-    0.065}^{+    0.036+    0.068}$ & $    0.771_{-    0.041-    0.069}^{+    0.041+    0.068}$ & $    0.768_{-    0.041-    0.064}^{+    0.041+    0.065}$  \\

$H_0$  & $   69.0_{-    1.7-    2.7}^{+    1.2+    3.1}$   & $   68.31_{-    0.82-    1.6}^{+    0.83+    1.6}$  & $   69.77_{-    0.71-    1.4}^{+    0.74+    1.4}$ & $   69.82_{-    0.75-    1.4}^{+    0.75+    1.5}$   \\

$S_8$  &  $    0.802_{-    0.020-    0.035}^{+    0.019+    0.037}$  & $    0.802_{-    0.019-    0.035}^{+    0.019+    0.037}$  & $    0.799_{-    0.020-    0.036}^{+    0.020+    0.037}$ & $    0.796_{-    0.018-    0.032}^{+    0.018+    0.033}$  \\

$r_{\rm{drag}}$  &  $  147.15_{-    0.30-    0.59}^{+    0.30+    0.60}$   & $  147.14_{-    0.30-    0.58}^{+    0.30+    0.60}$ & $  147.06_{-    0.29-    0.59}^{+    0.29+    0.56}$ & $  147.10_{-    0.29-    0.57}^{+    0.29+    0.56}$  \\

%\hline
%$\chi^2_{\rm bf}$ & $2777.43$ & $3813.08$ & $3825.82$ & $3832.84$ \\
%$\Delta \chi^2_{\rm bf}$ & $-2.28$ & $-1.1$ & $-4.5$ & $-7.57$ 
%\\

\hline\hline                                               
\end{tabular}          }                                     
\caption{68\% and 95\% CL constraints on several free and derived parameters of the interacting scenario \textbf{IDEp} ensuring the stability criterion using a variety of cosmological data sets.}
\label{tab:IDEpOK2}                                              
\end{table*}                                            
\end{center}                                                    
\endgroup

%%%%%%%%%%%%%%%%%%%%%%%%%%%%%%%%%%%%%%%%%%%%

\begin{figure*}
\includegraphics[width=0.9\textwidth]{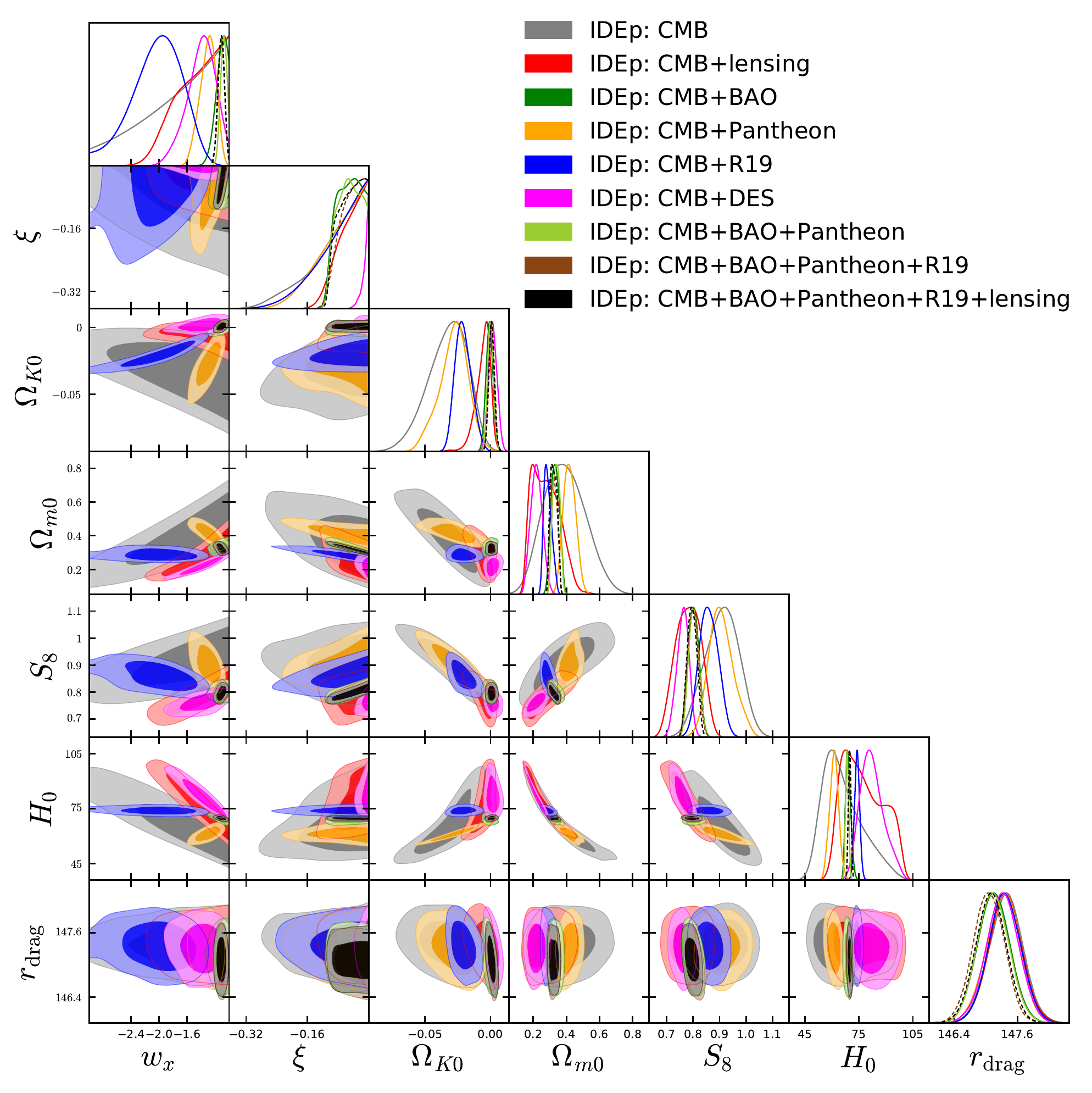}
\caption{One dimensional posterior distributions and two dimensional joint contours for the interacting scenario \textbf{IDEp} arising from the analyses of several cosmological data sets. }
\label{fig-phantom}
\end{figure*}

\begin{table}[h]
\centering
\begin{tabular}{|c|c|c|c|c|}
\hline 
Data & $\ln B_{ij}$\\
\hline 
CMB &  $6.6$\\
CMB+lensing & $1.9$ \\
CMB+Pantheon & $4.9$ \\
CMB+R19 & $12.2$\\
CMB+DES & $-1.2$ \\
CMB+BAO & $-3.4$\\
CMB+BAO+Pantheon & $-4.8$ \\
CMB+BAO+Pantheon+R19 & $-1.8$\\ 
CMB+BAO+Pantheon+R19+lensing & $0.6$ \\
\hline 
\end{tabular}
\caption{The table presents the $\ln B_{ij}$ values computed for the interacting scenario {\bf IDEp} with respect to the reference model $\Lambda$CDM considering all the datasets that are shown in Tables~\ref{tab:IDEpOK} and \ref{tab:IDEpOK2}. We note that a negative value  in $\ln B_{ij}$ obtained for a specific dataset denotes the preference for the $\Lambda$CDM model the over {\bf IDEp} scenario, 
while a positive value refers exactly to the opposite situation.    }
\label{tab:BE-IDEp}
\end{table}
%%%%%%%%%%%%%%%%%%%%%%%%%%%%%%%%%%%%%%%%%%%%

\subsection{IDEp: $\xi <0$, $w_x< -1$}

The results for the cosmological parameters within this phantom and interacting dark energy model are shown in  Tables~\ref{tab:IDEpOK} and~\ref{tab:IDEpOK2} and Fig.~\ref{fig-phantom} for the different observational data sets. 

From the analyses of the CMB data alone, as shown  in the second column of Tab.~\ref{tab:IDEpOK}, we find that the interaction is perfectly consistent with $\xi =0$ within $1\sigma$. The preference for a closed universe found in previous analyses in the literature still persists, and it is confirmed at more than two standard deviations. We report here a value of $\Omega_{K0} = -0.032_{-0.031}^{+0.028}$ at 95\% CL. 
Note that the Hubble constant has a very low mean value, $H_0 = 66_{-15}^{+  7}$~km/s/Mpc at 68\% CL. The very large error bars on $H_0$ allow to solve the tension with R19 within one standard deviation. In this case, the amelioration of the $H_0$ tension is mainly driven by an increase of the volume in the parameter space, rather than being due to an increase of the mean value of the Hubble constant. The reason for the lower value of $H_0$ is due to the strong anti-correlation between $\Omega_{m0}$ and $H_0$, see Fig.~\ref{fig-phantom}: the (negative) DM-DE interaction $\xi$, implies a flux of energy from the DE to the DM sector, producing a larger value of $\Omega_{m0}$ and consequently a lower value of $H_0$. The strong geometrical degeneracy between the DE equation of state, $w_x$, and $H_0$ relaxes the error bars on the Hubble constant. The DE equation of state is in agreement with the cosmological constant value $w_x=-1$ within $1\sigma$. Consequently, the CMB alone data  still prefer a closed universe. Also, the fact that $w_x$ is allowed to freely vary in the phantom region helps in alleviating the $H_0$ tension. However, if we focus on the estimated value of $S_8$,  this parameter has a higher value compared to the cosmic shear measurements. For instance KiDS-1000 analyses provide $S_8 =
0.766^{+0.020}_{-0.014}$ (at 68\% CL)~\cite{KiDS:2020suj} and therefore the tension on $S_8$ for CMB data alone is not alleviated.   

If we add CMB lensing, (third column of Tab.~\ref{tab:IDEpOK}), we still find a consistency with $\xi =0$, but the preference for a closed universe disappears ($\Omega_{K0} = -0.0058_{-0.0037}^{+0.0068}$ at 68\% CL). In this case the Hubble constant $H_0 = 77_{-16}^{+  9}$~km/s/Mpc at 68\% CL, and $S_8=0.782^{+0.052}_{-0.048}$ at 68\% CL, solving the tension with R19 and the weak lensing data\footnote{See~\cite{DiValentino:2020vvd} for a recent overview.} within one standard deviation, while the DE equation of state is in agreement with the cosmological constant.

When BAO observations are added to the CMB ones (second column of Tab.~\ref{tab:IDEpOK2}), we observe some changes in the constraints, due to their strong disagreement when the spatial curvature is a free parameter~\cite{DiValentino:2019qzk,Handley:2019tkm}. The curvature becomes perfectly consistent with a spatially flat universe (equivalently, $\Omega_{K0} = 0$), and there is a very mild preference both for a coupling in the dark sector,  $\xi=-0.052^{+0.045}_{-0.025}$ at 68\% CL, and for a phantom model, $w_x=-1.12^{+0.11}_{-0.03}$ at 68\% CL. In this case, because of the positive correlation between $\Omega_{K0}$ and $H_0$, see Fig.~\ref{fig-phantom}, the Hubble constant value is increased to $H_0 = 69.0_{- 1.7}^{+    1.2}$~ km/s/Mpc at 68\% CL, alleviating the tension with R19 at $2.3\sigma$ and the $S_8$ value is decreased ($S_8 = 0.802^{+0.019}_{-0.020}$ at 68\% CL for CMB+BAO) compared to the CMB alone, however, that does not help much when we look at the estimation from KiDS-1000 ($S_8 =
0.766^{+0.020}_{-0.014}$)~\cite{KiDS:2020suj}.  However, note that these two data sets, i.e. CMB and BAO, are in tension when considering closed cosmologies. 

A completely different result arises from the combination of the Pantheon Supernovae Ia compilation and CMB measurements: in this case, the preference for a closed universe is increased. We obtain $\Omega_{K0} = -0.028_{- 0.024}^{+      0.020}$ at 95\% CL, shifting the value of $H_0$ towards much lower values than those quoted above: $H_0 = 61.1_{-    2.5}^{+    2.7}$~km/s/Mpc at 68\% CL. The tension with R19 has a statistical significance of $4.2\sigma$. Also, the tension on the clustering related parameter $S_8$ gets worse.
While the combination of CMB+Pantheon data sets is perfectly in agreement with no interaction in the dark sector, a strong indication for a phantom universe appears at more than $2\sigma$,  $w_x=-1.31^{+0.22}_{-0.24}$ at 95\% CL.
Therefore, the inclusion Pantheon data to CMB observations increases both the Hubble tension and the preference for a phantom closed universe (see also Refs.~\cite{DiValentino:2020hov,Shirokov:2020dwl,Vagnozzi:2020dfn}).

We then move to the CMB+R19 combination shown in the fifth column of Tab.~\ref{tab:IDEpOK}. We note that here we can safely add these two data sets because they are not in tension. The addition of the $H_0$ prior will lead to a very negative value for the phantom DE equation of state, $w_x  = -2.03_{-0.30}^{+    0.43}$ at 68\% CL, ruling out the cosmological constant hypothesis with a high significance. This CMB+R19 combination also prefers a closed universe at more than 2 standard deviations, $\Omega_{K0} =   -0.021_{-0.012}^{+  0.013}$ at 95\% CL, but it does not prefer an interaction. Consequently, the CMB+R19 data set is completely in agreement with R19 at the price of a phantom closed universe. However, the tension on $S_8$ is not alleviated in this case.  

After that we also considered a combination of CMB and DES (see the last column of Table~\ref{tab:IDEpOK}) aiming to understand how the inclusion of galaxy clustering and cosmic shear measurements from the DES survey affects the constraints obtained from CMB alone (see the results in the second column of Table~\ref{tab:IDEpOK}). We find that the interaction is consistent with zero within  $1\sigma$ and $w_x$ is also consistent with the cosmological constant value within $1\sigma$. The indication of a curved universe that appeared in the CMB alone analysis disappears here ($\Omega_{K0} = 0.0017_{-0.0036}^{+    0.0031}$ at 68\% CL), meaning that a spatially flat universe  is supported by the combination of CMB+DES. However, we have some changes in the constraints on both $H_0$ and $S_8$.  We find that the Hubble constant increases in this case,  leading to a value of $H_0 =  82.3_{- 8.0}^{+    5.4}$~km/s/Mpc at 68\% CL, and hence the tension with R19 is alleviated within $1.5\sigma$. Similarly, we find that the estimation of $S_8$, $S_8 = 0.761_{-    0.021}^{+    0.026}$ at 68\% CL, is lower (if compared to the CMB case) and hence this resolves the $S_8$ tension within $1\sigma$.

Finally, we consider the last  three possible combinations of the data sets exploited here, CMB+BAO+Pantheon, CMB+BAO+Pantheon+R19, and CMB+BAO+Pantheon+R19+lensing, shown in the last 3 columns of Tab.~\ref{tab:IDEpOK2}. We find that for all these combinations, due to the presence of the BAO data, the preference for a phantom closed universe is diluted, the Hubble constant tension is not solved, and the preference for a dark sector coupling is either very mild or completely insignificant. 

We have also performed  the Bayesian evidence analyses for this interacting scenario with respect to the spatially flat $\Lambda$CDM model following the methodology as described at the end of Sec.~\ref{sec-data}.  We present the Bayes factors $\ln B_{ij}$ for this scenario in Table~\ref{tab:BE-IDEp} considering all the datasets and their combinations for which the constraints are reported (see Tables~\ref{tab:IDEpOK} and \ref{tab:IDEpOK2}). As described in Table~\ref{tab:BE-IDEp}, a negative value in  $\ln B_{ij}$ indicates a preference for the flat $\Lambda$CDM model over the {\bf IDEp} scenario, therefore, following this convention, we see that this interacting scenario is preferred over the flat $\Lambda$CDM for a number of observational datasets, such as CMB, CMB+lensing, CMB+Pantheon, CMB+R19 and the final combination of CMB+BAO+Pantheon+R19+lensing. Nevertheless, for some datasets, specially CMB+DES, CMB+BAO, CMB+BAO+Pantheon, and CMB+BAO+Pantheon+R19, the $\Lambda$CDM model is preferred over this interacting scenario.

%%%%%%%%%%%%%%%%%%%%%%%%%%%%%%%%%%%%%%%%%%%%
\begingroup                                                                                                                   \squeezetable                                        
\begin{center}                                                                                                                 
\begin{table*}                                                                                                                 \resizebox{\textwidth}{!}{  
\begin{tabular}{cccccccccc}                                                                                                            
\hline\hline                                                                                                                    
Parameters & CMB & CMB+lensing & CMB+Pantheon & CMB+R19 & CMB+DES   \\ \hline

$\Omega_{\rm c} h^2$ & $    0.064_{-    0.029-    0.058}^{+    0.044+    0.052}$ & $    0.072_{-    0.021-    0.065}^{+    0.047+    0.049}$ & $    0.036_{-    0.033}^{+    0.012}\,<0.074$ & $    <0.024\,<0.059$   & $    0.091_{-    0.013-    0.020}^{+    0.010+    0.023}$ \\

$\Omega_{\rm b} h^2$ & $    0.02262_{-    0.00017-    0.00032}^{+    0.00016+    0.00034}$ & $    0.02250_{-    0.00016-    0.00032}^{+    0.00016+    0.00033}$  & $    0.02258_{-    0.00017-    0.00032}^{+    0.00017+    0.00033}$ & $    0.02242_{-    0.00016-    0.00032}^{+    0.00016+    0.00033}$  & $    0.02248_{-    0.00016-    0.00033}^{+    0.00016+    0.00032}$ \\

$100\theta_{MC}$ & $    1.0447_{-    0.0033-    0.0039}^{+    0.0017+    0.0047}$ & $    1.0441_{-    0.0032-    0.0036}^{+    0.0013+    0.0051}$  & $    1.0467_{-    0.0018-    0.0034}^{+    0.0022+    0.0034}$ & $    1.0480_{-    0.0009-    0.0036}^{+    0.0021+    0.0027}$  & $    1.04265_{-    0.00073-    0.0015}^{+    0.00082+    0.0014}$  \\

$\tau$ & $    0.0474_{-    0.0074-    0.016}^{+    0.0081+    0.015}$ &  $    0.0487_{-    0.0077-    0.017}^{+    0.0078+    0.016}$  & $    0.0496_{-    0.0074-    0.016}^{+    0.0075+    0.015}$  & $    0.0540_{-    0.0082-    0.014}^{+    0.0071+    0.016}$  & $    0.0549_{-    0.0078-    0.016}^{+    0.0077+    0.016}$  \\

$n_s$ & $    0.9716_{-    0.0047-    0.0094}^{+    0.0047+    0.0092}$ &  $    0.9693_{-    0.0046-    0.0090}^{+    0.0046+    0.0092}$   & $    0.9703_{-    0.0045-    0.0093}^{+    0.0045+    0.0090}$ & $    0.9668_{-    0.0045-    0.0090}^{+    0.0044+    0.0086}$  & $    0.9678_{-    0.0048-    0.0089}^{+    0.0046+    0.0091}$  \\

${\rm{ln}}(10^{10} A_s)$ & $    3.025_{-    0.016-    0.033}^{+    0.017+    0.032}$ &  $    3.028_{-    0.016-    0.034}^{+    0.016+    0.033}$  & $    3.031_{-    0.015-    0.032}^{+    0.016+    0.031}$  & $    3.042_{-    0.017-    0.029}^{+    0.015+    0.032}$ & $    3.043_{-    0.016-    0.034}^{+    0.016+    0.033}$  \\

$w_x$ & $   -0.53_{-    0.35}^{+    0.25},\,unconstr.$ & $   <-0.74\,<-0.49$  & $  <-0.88\,<-0.81$ & $  <-0.96\,<-0.91$ & $ <-0.94 < -0.85 $  \\

$\xi$ & $    0.27_{-    0.14}^{+    0.17}\,<0.49$ &  $  <0.20\,<0.30$  & $    0.320_{-    0.078-    0.18}^{+    0.099+    0.16}$  & $    0.249_{-    0.024-    0.10}^{+    0.055+    0.08}$   & $    0.087_{-    0.039-    0.081}^{+    0.037+    0.072}$ \\

$\Omega_{K0}$ & $   -0.087_{-    0.021-    0.11}^{+    0.059+    0.08}$ &  $   -0.0180_{-    0.0061-    0.026}^{+    0.0014+    0.020}$   & $   -0.0300_{-    0.0098-    0.019}^{+    0.0097+    0.019}$ & $   -0.0025_{-    0.0036-    0.0072}^{+    0.0035+    0.0077}$  & $   -0.0052_{-    0.0061-    0.013}^{+    0.0068+    0.013}$  \\

$\Omega_{m0}$ & $    0.43_{-    0.30-    0.38}^{+    0.15+    0.44}$ &  $    0.28_{-    0.17-    0.24}^{+    0.11+    0.25}$  & $    0.160_{-    0.076-    0.10}^{+    0.045+    0.11}$ & $    0.083_{-    0.042-    0.052}^{+    0.013+    0.078}$  & $    0.259_{-    0.024-    0.042}^{+    0.020+    0.044}$ \\

$\sigma_8$ & $    1.26_{-    0.71-    0.9}^{+    0.16+    1.5}$ &  $    1.30_{-    0.71-    0.9}^{+    0.09+    1.7}$ & $    2.1_{-    1.0-    1.2}^{+    0.4+    1.7}$  & $    2.9_{-    1.0-    1.7}^{+    1.2+    1.7}$    & $    0.978_{-    0.089-    0.176}^{+    0.085+    0.175}$  \\

$H_0$ & $   47.3_{-    7.5-   13}^{+    7.1+   13}$ &  $   60.6_{-    5.2-    13}^{+    7.6+    11}$  & $   60.5_{-    2.5-    4.5}^{+    2.1+    4.7}$  & $   72.5_{-    1.4-    2.8}^{+    1.4+    2.7}$   & $   66.5_{-    2.3-    4.8}^{+    2.5+    4.5}$ \\

$S_8$ & $    1.27_{-    0.27-    0.35}^{+    0.09+    0.53}$ & $    1.06_{-    0.21-    0.26}^{+    0.04+    0.48}$  & $    1.38_{-    0.33-    0.42}^{+    0.18+    0.53}$  & $    1.40_{-   0.20-    0.46}^{+    0.32+    0.41}$   & $    0.905_{-    0.060-    0.11}^{+    0.052+    0.11}$ \\

$r_{\rm{drag}}$ & $  147.36_{-    0.31-    0.60}^{+    0.31+    0.60}$ & $  147.38_{-    0.30-    0.59}^{+    0.31+    0.60}$ & $  147.33_{-    0.30-    0.61}^{+    0.31+    0.59}$ & $  147.19_{-    0.30-    0.60}^{+    0.30+    0.58}$   & $  147.30_{-    0.32-    0.57}^{+    0.30+    0.60}$ \\

\hline\hline                                                                                                                    
\end{tabular} }
\caption{68\% and 95\% CL constraints on several free and derived parameters of the interacting scenario \textbf{IDEq} ensuring the stability criterion using a variety of cosmological datasets. }
\label{tab:IDEqOK}                                                                                                   
\end{table*}                                                                                                                     
\end{center}                                                                                                                    
\endgroup

%%%%%%%%%%%%%%%%%%%%%%%%%%%%%%%%%%%%%%%%%%%%

%%%%%%%%%%%%%%%%%%%%%%%%%%%%%%%%%%%%%%%%%%%%
\begingroup                                                                                                                   \squeezetable                                        
\begin{center}                                                                                                                 
\begin{table*}                                                                                                                 \resizebox{\textwidth}{!}{  
\begin{tabular}{cccccccccc}                                                                                                            
\hline\hline                                                                                                                    
Parameters  & CMB+BAO  & CMB+BAO & CMB+BAO & CMB+BAO \\ 
 &  & +Pantheon &+Pantheon+R19 &+Pantheon+R19+lensing \\ \hline

$\Omega_{\rm c} h^2$  & $    0.071_{-    0.022-    0.061}^{+    0.042+    0.050}$ & $  0.075_{-    0.017}^{+    0.039}\,<0.11$ & $  0.067_{-    0.021}^{+    0.035}\,<0.11$  & $  0.069_{-    0.018 - 0.056}^{+    0.033 +0.044}$ \\

$\Omega_{\rm b} h^2$  & $    0.02239_{-    0.00015-    0.00031}^{+    0.00016+    0.00030}$  & $    0.02239_{-    0.00015-    0.00030}^{+    0.00015+    0.00030}$ & $    0.02237_{-    0.00016-    0.00030}^{+    0.00016+    0.00032}$  & $    0.02239_{-    0.00015-    0.00029}^{+    0.00015+    0.00030}$  \\

$100\theta_{MC}$  & $    1.0441_{-    0.0030-    0.0036}^{+    0.0014+    0.0047}$  & $    1.0438_{-    0.0026-    0.0033}^{+    0.0010+    0.0049}$  & $    1.0443_{-    0.0026-    0.0034}^{+    0.0013+    0.0044}$ & $    1.0442_{-    0.0024-    0.0032}^{+    0.0011+    0.0043}$  \\

$\tau$  &  $    0.0547_{-    0.0082-    0.015}^{+    0.0073+    0.016}$  & $    0.0555_{-    0.0082-    0.016}^{+    0.0077+    0.016}$  & $    0.0548_{-    0.0078-    0.016}^{+    0.0078+    0.017}$ & $    0.0555_{-    0.0074-    0.014}^{+    0.0073+    0.015}$  \\

$n_s$  &  $    0.9660_{-    0.0044-    0.0086}^{+    0.0044+    0.0088}$   & $    0.9661_{-    0.0045-    0.0087}^{+    0.0045+    0.0086}$ & $    0.9650_{-    0.0046-    0.0092}^{+    0.0046+    0.0092}$ & $    0.9651_{-    0.0044-    0.0086}^{+    0.0044+    0.0087}$  \\

${\rm{ln}}(10^{10} A_s)$ &  $    3.044_{-    0.015-    0.030}^{+    0.015+    0.032}$   & $    3.046_{-    0.016-    0.032}^{+    0.016+    0.034}$  & $    3.046_{-    0.016-    0.033}^{+    0.016+    0.035}$ & $    3.047_{-    0.015-    0.028}^{+    0.014+    0.030}$  \\

$w_x$  & $   <-0.88\,<-0.78$  & $  <-0.88\,<-0.78$  & $  <-0.90\,<-0.82$  & $  <-0.91\,<-0.83$  \\

$\xi$  &  $    0.13_{-    0.10}^{+    0.06}\,<0.26$   &  $    0.119_{-    0.095}^{+    0.048}\,<0.25$ & $    0.139_{-    0.077-    0.11}^{+    0.059+    0.12}$ & $    0.135_{-    0.076-    0.10}^{+    0.051+    0.13}$   \\

$\Omega_{K0}$  &  $   0.0000_{-    0.0030-    0.0057}^{+    0.0025+    0.0058}$   & $    0.0000_{-    0.0022-    0.0043}^{+    0.0022+    0.0044}$  & $    0.0010_{-    0.0021-    0.0042}^{+    0.0021+    0.0045}$ & $    0.0008_{-    0.0021-    0.0041}^{+    0.0021+    0.0044}$ \\

$\Omega_{m0}$  &  $    0.202_{-    0.056-    0.14}^{+    0.095+    0.12}$  & $    0.212_{-    0.040-    0.14}^{+    0.085+    0.10}$  & $    0.186_{-    0.047-    0.12}^{+    0.073+    0.10}$ & $    0.189_{-    0.041-    0.12}^{+    0.070+    0.09}$  \\

$\sigma_8$ &  $    1.36_{-    0.65-    0.8}^{+    0.11+    1.5}$  & $    1.27_{-    0.54-    0.7}^{+    0.62+    1.4}$   & $    1.39_{-    0.58-    0.8}^{+    0.11+    1.3}$  & $    1.35_{-    0.52-    0.7}^{+    0.08+    1.2}$   \\

$H_0$  &  $   68.5_{-    1.4-    2.7}^{+    1.4+    2.8}$  & $   68.24_{-    0.84-    1.6}^{+    0.85+    1.7}$  & $   69.72_{-    0.75-    1.5}^{+    0.75+    1.4}$ & $   69.75_{-    0.74-    1.4}^{+    0.74+    1.5}$  \\

$S_8$  & $    1.00_{-    0.20-    0.25}^{+    0.05+    0.43}$  & $    0.98_{-   0.17-    0.22}^{+    0.03+    0.42}$  & $    1.01_{-   0.19-    0.24}^{+    0.05+    0.39}$ & $    1.00_{-   0.17-    0.22}^{+    0.04+    0.37}$   \\

$r_{\rm{drag}}$  & $  147.14_{-    0.30-    0.60}^{+    0.30+    0.60}$ & $  147.16_{-    0.30-    0.60}^{+    0.30+    0.58}$  & $  147.07_{-    0.30-    0.58}^{+    0.30+    0.58}$  & $  147.06_{-    0.29-    0.56}^{+    0.29+    0.57}$  \\

\hline\hline                                                                                                                    
\end{tabular} }
\caption{68\% and 95\% CL constraints on several free and derived parameters of the interacting scenario \textbf{IDEq} ensuring the stability criterion using a variety of cosmological datasets. }
\label{tab:IDEqOK2}                                                                                                   
\end{table*}                                                                                                                     
\end{center}                                                                                                                    
\endgroup   

%%%%%%%%%%%%%%%%%%%%%%%%%%%%%%%%%%%%%%%%%%%%

\begin{figure*}
\includegraphics[width=0.9\textwidth]{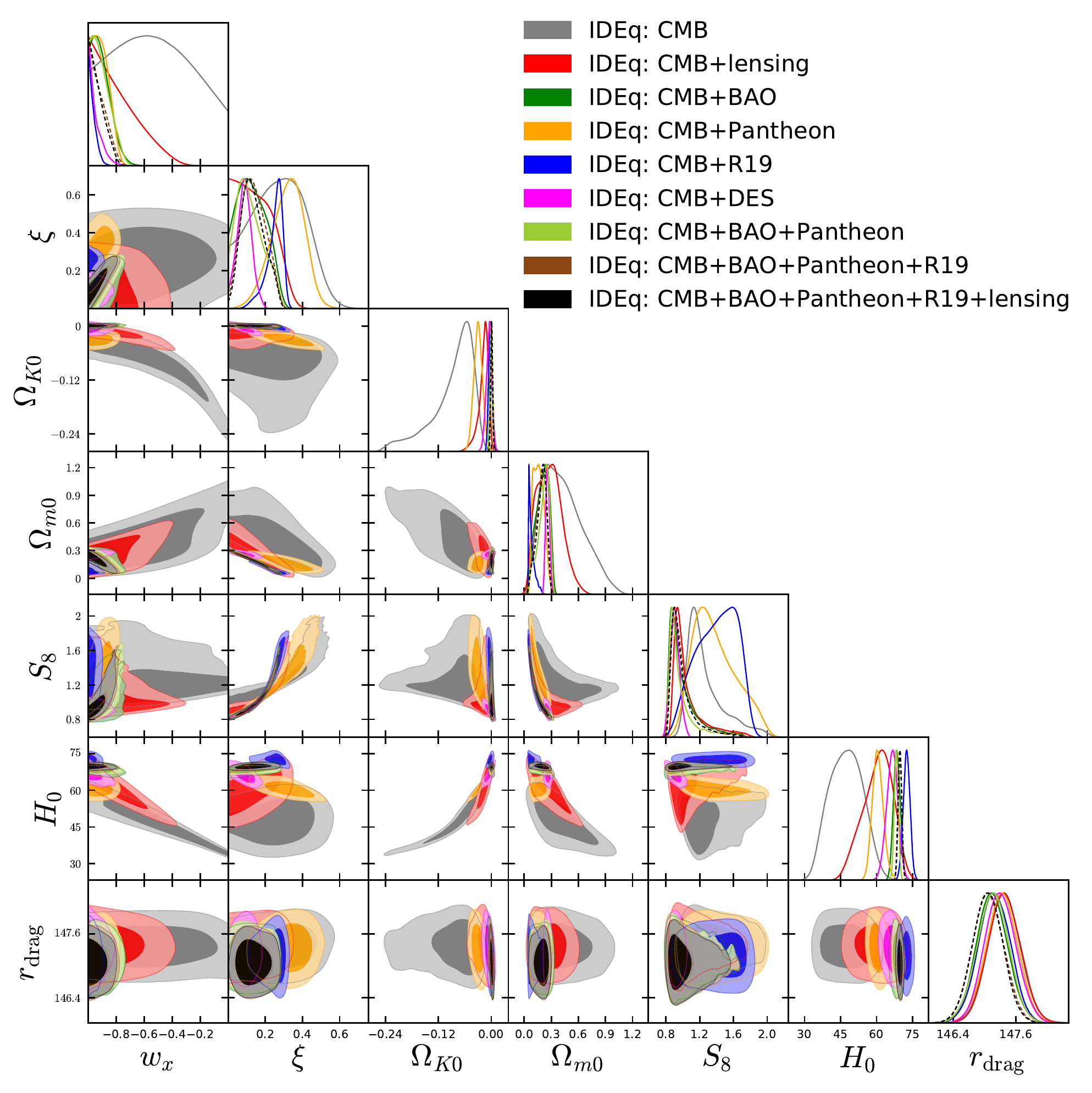}
\caption{As Fig.~\ref{fig-phantom} but for the interacting scenario \textbf{IDEq}. }
\label{fig-quin}
\end{figure*}
\begin{table}[h]
\centering
\begin{tabular}{|c|c|c|c|c|}
\hline 
Data & $\ln B_{ij}$\\
\hline 
CMB &  $5.9$\\
CMB+lensing & $1.6$ \\
CMB+Pantheon & $1.9$ \\
CMB+R19 & $4.2$\\
CMB+DES & $-2.5$ \\
CMB+BAO & $-4.4$\\
CMB+BAO+Pantheon & $-4.9$ \\
CMB+BAO+Pantheon+R19 & $-3.0$\\ 
CMB+BAO+Pantheon+R19+lensing & $0.5$ \\
\hline 
\end{tabular}
\caption{The table presents the $\ln B_{ij}$ values computed for the interacting scenario {\bf IDEq} with respect to the reference model $\Lambda$CDM considering all the datasets that are shown in Tables~\ref{tab:IDEqOK} and \ref{tab:IDEqOK2}. We note that a negative value  in $\ln B_{ij}$ obtained for a specific dataset denotes the preference for the  $\Lambda$CDM model over the {\bf IDEq} scenario, while a positive value indicates exactly the opposite situation.   }
\label{tab:BE-IDEq}
\end{table}
%%%%%%%%%%%%%%%%%%%%%%%%%%%%%%%%%%%%%%%%%%%%

\subsection{IDEq: $\xi >0$, $w_x> -1$}

The results for the quintessential scenario are summarized in Tables~\ref{tab:IDEqOK} and~\ref{tab:IDEqOK2} and Fig.~\ref{fig-quin} considering CMB from Planck 2018 and its combination with other external data sets. 

As before, we shall start by discussing the results obtained from CMB data alone, which are shown in the second column of Tab.~\ref{tab:IDEqOK}. 
Notice that there is a mild preference for a non-zero dark sector interaction at one standard deviation, $\xi=0.27^{+0.17}_{-0.14}$ at 68\% CL. The preference for a closed universe is stronger,  surpassing the $2\sigma$ significance, $\Omega_{K0} = -0.09_{-  0.11}^{+0.08}$ at 95\% CL. However, the $H_0$ tension with R19 is above $3\sigma$, because we obtain a value of the Hubble constant of $H_0 = 47.3_{- 7.5}^{+7.1}$~km/s/Mpc at 68\% CL, even if $H_0$ has very large error bars. In this case, the strong negative degeneracy between $w_x$ and $H_0$ leads to very low values for $H_0$, due to the mean preferred value of the DE equation of state,  $w_x=-0.53^{+0.25}_{-0.35}$ at 68\% CL, which differs from the cosmological constant picture at more than $1\sigma$. CMB alone data prefers a quintessential interacting closed universe, and the tension with R19 is quite significant. Lastly, the $S_8$ parameter takes a very large value ($S_8 = 1.27_{-    0.27}^{+    0.09}$ at 68\% CL) compared to KiDS-1000  where $S_8 = 0.766^{+0.020}_{-0.014}$~\cite{KiDS:2020suj}, hence, the tension is clearly not alleviated at all.  

If we add the CMB lensing (third column of Tab.~\ref{tab:IDEqOK}), we still find a consistency with $\xi =0$, and an indication for a closed universe at $1\sigma$ ($\Omega_{K0} = -0.0180_{-0.0061}^{+0.0014}$ at 68\% CL). In this case the $H_0$ tension with R19 is within $3\sigma$, because we have $H_0 = 60.6_{- 13}^{+11}$~km/s/Mpc at 95\% CL, with very large error bars, and the DE equation of state is in agreement with the cosmological constant within $1\sigma$. However, about the $S_8$ parameter, our conclusion remains the same as in the CMB alone case. 

The inclusion of BAO observations  (second column of Tab.~\ref{tab:IDEqOK2}) pushes the Hubble constant to higher values, $H_0 = 68.5_{-1.4}^{+    1.4}$ km/s/Mpc at 68\% CL, reducing the tension with R19 to $2.8\sigma$. For CMB+BAO, the curvature parameter becomes perfectly consistent with $\Omega_{K0} = 0$, there is only a mild indication for the presence of a coupling in the dark sector ($\xi=0.13^{+0.06}_{-0.10}$ at 68\% CL), and the DE equation of state is in perfect agreement with the cosmological constant case, $w_x<-0.88$ at 68\% CL.  Also, we note that the tension on $S_8$ parameter is not alleviated.  

In analogy to the phantom closed interacting scenario, completely different results are obtained when combining the Pantheon compilation data with the CMB observations. In this case, we find again a larger preference for a closed universe, $\Omega_{K0} = -0.030\pm 0.019$ at 95\% CL, and the value of $H_0$ is also shifted towards lower values $H_0 = 60.5_{-    2.5}^{+    2.1}$~km/s/Mpc at 68\% CL, in strong disagreement with R19. However, in this quintessence scenario, the Planck+Pantheon combination leads to an indication for an interaction at more than $2\sigma$, $\xi=0.32^{+0.16}_{-0.18}$ at 95\% CL, and to a DE equation of state in perfect agreement with the cosmological constant scenario. Therefore, the  CMB+Pantheon data set is in disagreement with R19, but prefers a closed and interacting universe.  This combination also fails to alleviate the $S_8$ tension. 

We have also considered the combined analysis CMB+DES and  the results are presented in the last column of  Table~\ref{tab:IDEqOK}. Notice that the preference for a closed universe that we obtained for CMB alone disappears in this case and the spatially flatness of the universe is indicated within $1\sigma$.  However, we find a preference of a non-zero interaction at more than $2\sigma$ ($\xi = 0.087_{-    0.081}^{+0.072}$ at 95\% CL) while the dark energy equation of state, $w_x$, becomes consistent with the cosmological constant value. 
The Hubble constant increases, leading to $H_0 =  66.5_{- 2.3}^{+    2.5}$~km/s/Mpc at 68\% CL and due to these large error bars, the tension with R19 is reduced down to $2.6\sigma$. However, the $S_8$ tension is not alleviated in this case, even though we notice that $S_8$ is significantly lowered compared to CMB alone.  

The R19 data set is added in two different combinations and the results are shown in the fifth column of Tab.~\ref{tab:IDEqOK} and the last two columns of Tab.~\ref{tab:IDEqOK2} for completeness, but notice that these results should be regarded as not fully reliable due to the strong tension among CMB and R19 in the $w_x>-1$ region. Concerning the $S_8$ parameter, its tension is not alleviated for these cases. This is clear if one compares the values obtained in CMB+R19, CMB+BAO+Pantheon+R19, CMB+BAO+Pantheon+R19+lensing and KiDS-1000 \cite{KiDS:2020suj}.

Finally, the CMB+BAO+Pantheon combination provides a preference for a flat universe with $w_x=-1$ and a very mild indication for the presence of an interaction in the dark sector. Concerning the  $H_0$ value, there is still a disagreement with R19 with a significance of $3.5\sigma$. However, the tension on $S_8$ is not alleviated as one can see that the estimated value of $S_8$ for CMB+BAO+Pantheon is quite large compared to KiDS-1000 ($S_8 = 0.766^{+0.020}_{-0.014}$)~\cite{KiDS:2020suj}.

Lastly, we perform the Bayesian evidence analysis for this interacting scenario (with respect to the spatially flat $\Lambda$CDM model) considering all the datasets as shown in Tables~\ref{tab:IDEqOK} and \ref{tab:IDEqOK2}, and present the values of $\ln B_{ij}$ in Table~\ref{tab:BE-IDEq}. Since according to our convention, a negative value in  $\ln B_{ij}$ corresponds to the preference of the flat $\Lambda$CDM model over the {\bf IDEp} scenario and the positive value  in  $\ln B_{ij}$ refers exactly the opposite scenario, therefore, following this we see that for CMB, CMB+lensing, CMB+Pantheon, CMB+R19 and the final combination CMB+BAO+Pantheon+R19+lensing, the {\bf IDEq} is preferred over the flat $\Lambda$CDM. However, for some datasets, especially CMB+DES, CMB+BAO, CMB+BAO+Pantheon, and CMB+BAO+Pantheon+R19, $\Lambda$CDM is preferred over {\bf IDEq}.   

\section{Concluding remarks}
\label{sec-conclusion}

Cosmologies with non-gravitational interactions between dark matter and dark energy have been extensively studied in the literature for their ability to address the Hubble constant tension $-$ a discrepancy in the estimation of $H_0$ between  high and low redshift measurements. These interacting scenarios are very general, since they allow for an energy exchange mechanism between the dark components. A further generalization can be realized if the flatness condition is relaxed. This possibility is strongly motivated by the claimed preference for a closed universe in a large number of recent and independent analyses in the literature~\cite{Aghanim:2018eyx,Handley:2019tkm,DiValentino:2019qzk,DiValentino:2020hov}. 
Whether or not non-flat interacting cosmologies can alleviate the $H_0$ tension has been discussed before, see Ref.~\cite{DiValentino:2020kpf}. However, the framework was not as general and complete as possible since the dark energy equation of state was supposed to mimic the vacuum energy case and in interacting cosmologies this assumption  may no longer be valid.  
In the current article, we have allowed the dark energy equation of state to freely vary. To ensure the stability of the interacting models, we have separately discussed two distinct cases where the dark energy equation of state lies in the phantom region ($w_x< -1$, \textbf{IDEp}) or in the quintessence region ($w_x> -1$, \textbf{IDEq}). Both scenarios have been confronted against the latest cosmological observations, and the results have been summarized in Tab.~\ref{tab:IDEpOK} and Fig. \ref{fig-phantom} and in Tab.~\ref{tab:IDEqOK} and Fig. \ref{fig-quin} for the \textbf{IDEp} and the \textbf{IDEq} scenarios, respectively.

While a phantom closed universe can provide a compelling solution to the $H_0$ tension, because the CMB alone data give a Hubble constant value in agreement with R19 without the necessity of a coupling between the dark matter and dark energy sectors (see also~\cite{DiValentino:2020hov}), the same model is in agreement with Pantheon, but in strong disagreement with Baryon Acoustic Oscillation data. However, the agreement between Planck and the luminosity distance measurements is at the price of a tension between these different observables, namely CMB+Pantheon and CMB+R19. 
Future independent estimates of the Hubble constant, as, for instance, those from gravitational-wave standard siren measurements~\cite{Schutz:1986gp,Holz:2005df,Chen:2017rfc,DiValentino:2018jbh,Palmese:2019ehe} may shed light on the current, $2021$ cosmological tensions.

\section*{Acknowledgements}
The authors thank the referee for some essential comments that improved the quality of the article.
WY was supported by the National Natural Science Foundation of China under Grants No. 12175096 and No. 11705079, and Liaoning Revitalization Talents Program under Grant no. XLYC1907098.
SP gratefully acknowledges the Science and Engineering Research Board, Govt. of India, for their Mathematical Research Impact-Centric Support Scheme (File No. MTR/2018/000940).
EDV acknowledges the support of the Addison-Wheeler Fellowship awarded by the Institute of Advanced Study at Durham University.
OM is supported by the Spanish grants FPA2017-85985-P, PROMETEO/2019/083 and by the European ITN project HIDDeN (H2020-MSCA-ITN-2019//860881-HIDDeN).

%%%%%%%%%%%%%%%%%%%%%%%%%%%%%%%%%%%%%%%%%%%%%%%%%%%%%
\bibliography{biblio}

\end{document}